# Exploring Dual-Iron Atomic Catalysts for Efficient Nitrogen Reduction: A Comprehensive Study on Structural and Electronic Optimization

Zhe Zhang[a], Wenxin Ma[a], Jiajie Qiao[a], Xiaoliang Wu[a], Shaowen Yu[a], Weiye Hou[a], Xiang Huang[a], Rubin Huo[a], Hongbo Wu[*,b,c] and Yusong Tu[*,a]

[a]College of Physics Science and Technology, Yangzhou University, Jiangsu 225009, China

[b]School of Science, Yangzhou Polytechnic Institute, Yangzhou 225127, China

[c]College of Physics, Hebei Normal University, Shijiazhuang 050024, China

[*]Corresponding Authors: wuhb@ypi.edu.cn; ystu@yzu.edu.cn

**Abstract**

The nitrogen reduction reaction (NRR), as an efficient and green pathway for ammonia synthesis, plays a crucial role in achieving on-demand ammonia production. This study proposes a novel design concept based on dual-iron atomic sites and nitrogen-boron co-doped graphene ($Fe_2N_xB_y@G$) catalysts, exploring their high efficiency in NRR. By modulating the $N_xB_y$ co-doped ratios, we found that $Fe_2N_3B@G$ catalyst exhibited significant activity in the adsorption and hydrogenation of $N_2$ molecules, especially with the lowest free energy (0.32 eV) on NRR distal pathway, showing its excellent nitrogen activation capability and NRR performance. The computed electron localization function, crystal orbital Hamiltonian population, electrostatic potential map revealed that the improved NRR kinetics of $Fe_2N_3B@G$ catalyst derived by $N_3B$ co-doping induced optimization of Fe-Fe electronic environment, regulation of Fe-N bond strength, and the continuous electronic support during the $N_2$ breakage and hydrogenation. In particular, machine learning molecular dynamics (MLMD) simulations were employed to verify the high activity of $Fe_2N_3B@G$ catalyst in NRR, which reveal that $Fe_2N_3B@G$ effectively regulates the electron density of Fe-N bond, ensuring the smooth generation and desorption of $NH_3$ molecules and avoiding the competition with hydrogen evolution reaction (HER). Furthermore, the determined higher HER overpotential of $Fe_2N_3B@G$ catalyst can effectively inhibit the HER and enhance the selectivity toward NRR. In addition, $Fe_2N_3B@G$ catalyst also showed good thermal stability by MD simulations up to 500 K, offering its feasibility in practical applications. This study demonstrates the superior performance of $Fe_2N_3B@G$ in nitrogen reduction catalysis, and provides theoretical guidance for atomic catalyst design by the co-doping strategy and in-deep electronic environment modulation.

# 1. Introduction

Ammonia is an important chemical for fertilizer production and a promising energy carrier, which plays a vital role in agriculture and industry.[1] Electrocatalytic nitrogen reduction reaction (NRR) is expected to directly convert nitrogen and electricity generation by renewable energy into ammonia. NRR has the advantages of low energy consumption and low carbon dioxide emissions.[2, 3] Therefore, it is considered to be a promising technology that can alleviate energy and environmental problems.[4, 5] However, due to the lack of dipole moment and low polarizability of nitrogen molecules and the existence of competitive hydrogen evolution reaction (HER), the efficiency of electrocatalytic ammonia synthesis process is still limited by high overpotential, low current density and low selectivity.[6, 7] Transition metal-based dual-atom catalysts provide a new design strategy for ammonia synthesis reaction due to their low coordination structure and high atomic utilization.[8, 9] Compared with pure metal catalysts, dual-atom catalysts are usually composed of transition metal atoms and support. The special coordination structure enables them to effectively inhibit the occurrence of competitive HER, thus achieving a lower overpotential. Single-atom catalysts have similarities with dual-atom catalysts in coordination environment,[10-12] but the advantage of dual-atom catalysts is that they have flexible reaction sites and can effectively overcome the thermodynamic and kinetic barriers of multi-step ammonia synthesis reaction. Although great progress has been made in the design of dual-atom catalysts for NRR, both theoretically and experimentally dual-atom catalysts still face the challenge of simultaneously improving the Faraday efficiency of ammonia synthesis and the ammonia yield.

Recently, theoretical calculations and experiments have screened a variety of nitrogen reduction catalyst, including metal sulfides, metal nitrides, graphene, porous carbon materials, etc.[9, 13-15] Meanwhile, a variety of dual-atom catalysts based on precious metals and non-precious metals have also been developed for NRR. The active centers of these catalysts are mainly composed of metal atoms coordinated with nitrogen atoms or oxygen vacancies. When the constituent elements of the catalyst are fixed, the electronic structure of this material can be changed by adjusting the atomic distribution of the catalyst, designing vacancies, and doping atoms.[16-19] The changes in the electronic structure will cause redistribution of electrons and changes in the charge transformation, which will affect the activation of nitrogen on the dual-atom catalyst. Therefore, exploring the electronic optimization of the catalyst plays a vital role in improving the efficiency of ammonia synthesis.

Recent studies have shown that introducing bimetallic sites to regulate the adsorption properties of target intermediates is an effective method to improve the activity of NRRs. Meanwhile, the activity and selectivity of the catalyst can be optimized by using different transition metals and ligands. Nitrogen-doped graphene is a good transition metal support due to its atomic defects. Its perfect electronic properties can effectively change the substrate's adsorption of various intermediates.[20, 21] In addition, transition metal dimer-supported nitrogen-doped phthalocyanine and graphene have been shown high activity for NRR. Due to the synergistic effect between metals, these dual-atom catalysts show stronger ability to inhibit HER and enhance NRR activity than corresponding single-atom catalysts.[22-24] Furthermore, studies have shown that transition metal atoms with low valence states can more effectively activate nitrogen molecules. Zhang et al. demonstrated that dual-atom catalysts with two homonuclear transition metals and pyrrolic nitrogen atoms achieve highly efficient NRRs by suppressing hydrogen evolution.[25] Dual-atom catalysts have significant advantages in maintaining the low valence states of transition metals. Two adjacent transition metal atoms can be designed as a whole and coordinated with the atoms on the support.[26, 27] Due to the similar bonding environment with single-atom catalysts, the average charge transfer amount of each transition metal in dual-atom catalysts will be smaller, thus showing a relatively low valence state. In recent years, the doped graphene materials have become ideal supports for catalysts due to their excellent conductivity and chemical stability.[28, 29] Especially in dual-sites atomic catalysts, doping elements can effectively adjust the electronic environment of metal active center and further improve the catalytic performance.

In this work, we proposes a new NRR catalyst design scheme based on dual-Fe atom and boron nitrogen co-doped graphene ($Fe_2N_xB_y$@G). By adjusting the $N_xB_y$ co-doping ratios and regulating the electronic coordination environments of dual-Fe active center, we found that $Fe_2N_3B$@G catalyst exhibited the lowest free energy (0.32 eV) in NRR distal pathway, demonstrating its excellent nitrogen activation capability and high NRR activity. To deeply understand the reaction mechanism of $Fe_2N_3B$@G catalyst, we carried out investigation on its electron localization function, crystal orbital Hamilton population, electrostatic potential map, as well as performing machine learning molecular dynamics simulations. By comprehensive study on the structural, electrical, charge and bonding characteristics, we revealed how the $N_3B$ co-doped strategy regulate the electronic environment around dual-Fe atoms, optimize the electron supply capacity of dual-Fe sites during NRR, and improve the adsorption and activation ability of $N_2$ molecules. These results show that $Fe_2N_3B$@G catalyst can effectively promote the breakage and hydrogenation of $N_2$ molecules at lower energy barriers, and successfully suppress the competition due to HER process. Machine learning molecular dynamics simulation further verified the high efficiency of $Fe_2N_3B$@G catalyst through evaluating itsNRR dynamic processfrom nitrogen activation to ammonia generation.

## 2. Computational methods

The spin-polarized calculations were performed using the Vienna ab initio Simulation Package (VASP),[30, 31] based on density functional theory (DFT) within the projector-augmented wave (PAW) method.[32, 33] The exchange-correlation energy was characterized through the generalized gradient approximation (GGA) with the Perdew-Burke-Ernzerhof (PBE) exchange functional,[34] and the van der Waals (vdW) interactions were considered with the semi-empirical DFT-D3 method. The energy cutoff for the plane wave basis was set to 500 eV. The convergence criteria for energy was $10^{-5}$ eV, and the force on each atom was optimized less than 0.01 eV/Å. The Gamma-centered 3×3×1 $k$-point mesh in the Brillouin zone and a 4×3 supercell containing 72 carbon atoms were chosen for structural relaxation. The perpendicular vacuum region was set to 20 Å to avoid the interlayer interaction. The climbing image nudged elastic band (CI-NEB) method was used to determine the minimum energy pathways and corresponding energy barriers.[35] Furthermore, the Bader charge analysis was employed to calculate the amount of charge transfer. The VASPKIT code and VESTA software were used for data processing and graphics production.

For the exploration of the NRR on $Fe_2N_3B$@G catalyst, machine learning molecular dynamics (MLMD) simulations were applied. The machine-learned force field (MLFF) approach was employed to efficiently capture atomic interactions during the simulation, offering an alternative to traditional methods by significantly expanding the simulation's time and length scales while maintaining accuracy.[36, 37] The MLFF was trained on a dataset generated from DFT calculations of various catalyst configurations, including different nitrogen intermediate adsorption modes. Once trained, the MLFF allowed for efficient updates of atomic positions and velocities, guiding the MD simulations.[38, 39] The temperature was controlled using the Nosé-Hoover thermostat, and data sampling started after the system reached equilibrium.[40] The accuracy of the MLFF was confirmed by low energy root-mean-square errors (RMSE), ensuring reliable atomic force predictions. This approach enabled efficient simulation of the dynamic processes involved in nitrogen activation, hydrogenation, and ammonia desorption, providing valuable insights into the catalytic mechanisms at a molecular level.

## 3. Results and discussions

### 3.1 Design scheme of Fe$_2$N$_x$B$_y$@G atomic catalyst toward NRR

We first explored the structural design of Fe$_2$N$_x$B$_y$@G catalyst and its performance in the NRR. Through structural optimization, a catalyst model with Fe dual-atom embedded in boron-nitrogen co-doped graphene was designed, as shown in Figure 1(a). Boron-nitrogen co-doped graphene not only provides good electronic conductivity and stability, but also improves the catalyst's ability to adsorb and activate nitrogen molecules by adjusting the electronic environment of Fe atoms. The dual-Fe atoms form active centers on the catalyst surface, which can effectively promote the reduction reaction of nitrogen molecules at low energy. Figure 1(b) and Figure S1 further show the different B:N doping ratios (1:3, 2:2, 3:1, 4:0). The variation of ratios directly affects the electronic structure of the catalyst, thereby adjusting the adsorption mode and reaction pathway of nitrogen molecules. With the change of the B:N ratio, the electronic properties of the catalyst and the adsorption characteristics of nitrogen molecule also changed, significantly affecting the activity and selectivity of NRR. Finally, Figure 1(c) shows the four pathways of nitrogen reduction: Distal, Alternating, Consecutive and Enzymatic. These four pathways represent different mechanisms of nitrogen molecule reduction on the catalyst surface. By calculating the free energy of these pathways, the feasibility and selectivity of different pathways in NRR can be further evaluated. Figure 1 provides a theoretical framework for subsequent free energy analysis by showing the structural characteristics of the catalyst and different NRR pathways. The comparison of these structures and pathways not only reveals the potential performance of the catalyst in the nitrogen reduction process, but also provides new optimization ideas for atomic catalyst design.

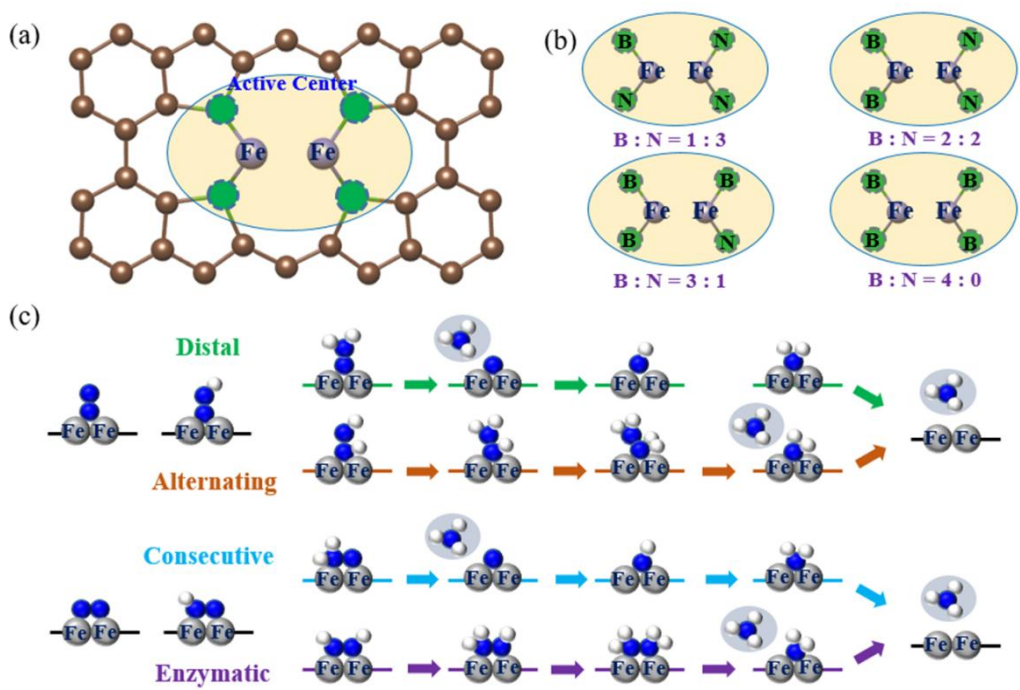

**Figure 1.** (a) Schematic diagrams of the crystal structures of the dual-Fe atomic catalyst anchored on boron-nitrogen doped graphene. (b) The doping structure with different ratios of B *vs* N. (c) Schematic diagrams of the four different pathways of NRR, including distal, alternating, consecutive and enzymatic pathways.

3.2 Catalytic performance of $Fe_2N_xB_y@G$ atomic catalyst toward NRR

Figure 2 shows the free energy changes of four catalysts $Fe_2N_3B@G$, $Fe_2N_2B_2@G$, $Fe_2NB_3@G$, and $Fe_2B_4@G$) during the NRR, focusing on catalytic performance of four reaction pathways. From the data in Figure 2(a), the free energy of $Fe_2N_3B@G$ catalyst on the distal pathway is the lowest (0.32 eV), which indicates that this pathway has the highest reaction activity, the smallest reaction energy barrier, and the nitrogen reduction process is most likely to occur. In contrast, the distal pathways of $Fe_2N_2B_2@G$ and $Fe_2NB_3@G$ catalysts cannot maintain the end-on structure due to the instability during nitrogen adsorption, and will eventually be optimized into a side-on structure, hence the distal pathway on these two catalysts are not existed. For $Fe_2N_2B_2@G$ and $Fe_2NB_3@G$, these two catalysts only have consecutive and enzymatic pathways, and their free energies are 0.61 eV ($Fe_2N_2B_2@G$) and 0.66 eV ($Fe_2NB_3@G$), respectively. These results show that the catalytic efficiencies of $Fe_2N_2B_2@G$ and $Fe_2NB_3@G$ on these pathways are relatively similar, and both show relatively mild reaction energy barriers, but are still lower than $Fe_2N_3B@G$.

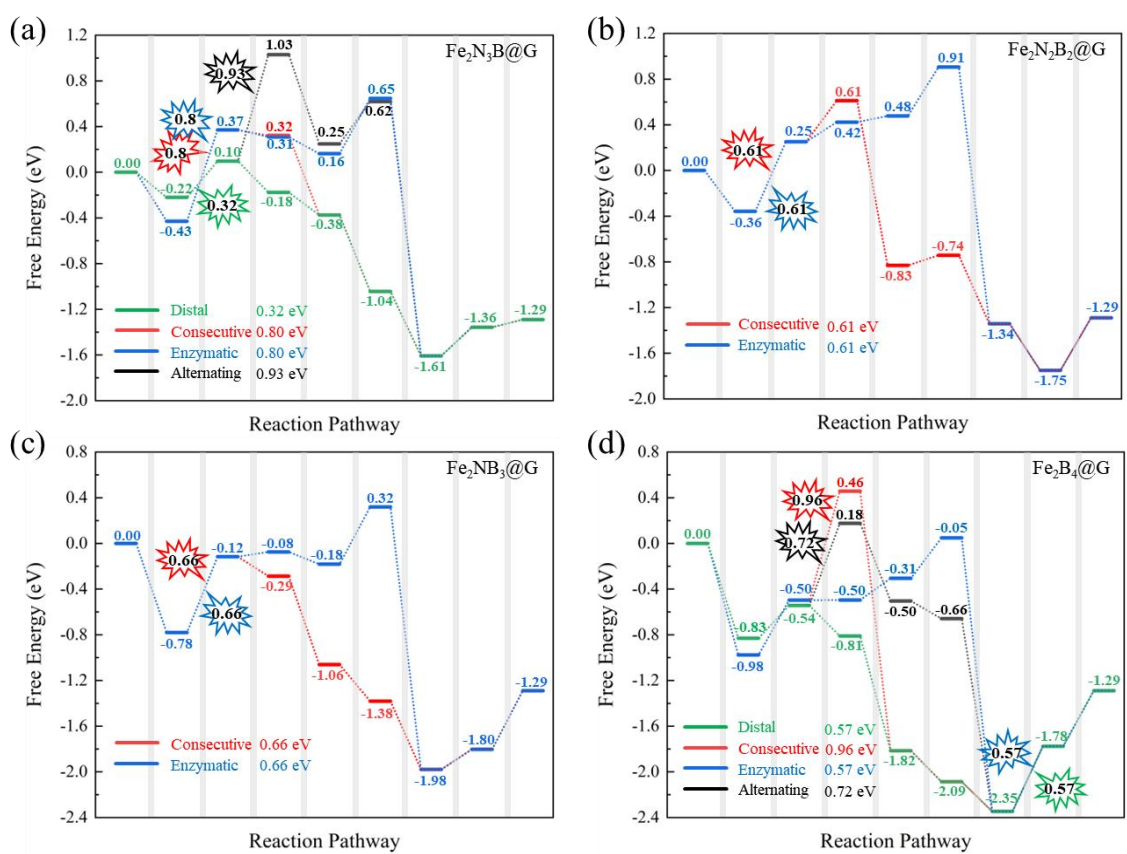

**Figure 2.** Free energy changes of the distal, alternating, consecutive and enzymatic pathways of NRR on catalysts of (a) $Fe_2N_3B@G$; (b) $Fe_2N_3B_2@G$; (c) $Fe_2NB_3@G$; (d) $Fe_2B_4@G$.

As seen in Figure 2(a), our computed results show that the free energies of the $Fe_2N_3B@G$ catalyst on the consecutive and enzymatic pathways are 0.80 eV and 0.80 eV, respectively, showing lower reactivity than the distal pathway. The free energy of the $Fe_2B_4@G$ catalyst is higher, with free energies of 0.96 eV and 0.72 eV on the consecutive and alternating pathways, respectively, indicating that its catalytic efficiency is low and the energy barrier is large during the reaction. In summary, the free energy analysis verifies that $Fe_2N_3B@G$ performs best among four $Fe_2N_xB_y@G$ catalysts and all NRR pathways, and its lower free energy of 0.32 eV on the distal pathway makes it a catalytically active NRR catalyst. Although $Fe_2N_2B_2@G$ and $Fe_2NB_3@G$ do not have distal pathways, they still have good catalytic performance on the consecutive and enzymatic pathways. Figure S2 further corroborates these findings by showing the optimal free energy pathways for each catalyst under different applied potentials. Specifically, the $Fe_2N_3B@G$ catalyst shows the lowest energy profile at -0.32V for the distal pathway, aligning with the favorable results observed in Figure 2(a). In comparison, the higher applied potentials required for $Fe_2N_2B_2@G$ and $Fe_2NB_3@G$ under consecutive and enzymatic pathways are reflected by the higher energy values in both Figures 2(b-c), consistent with their less efficient NRR performance. Through these free energy comparisons, we can provide a clearer idea for the optimization design of dual-Fe catalyst, especially by adjusting the metal coordination and $N_xB_y$ co-doping ratio to improve the efficiency toward NRR.

### 3.3 Electronic and charge properties of Fe$_2$N$_3$B@G atomic catalyst toward NRR

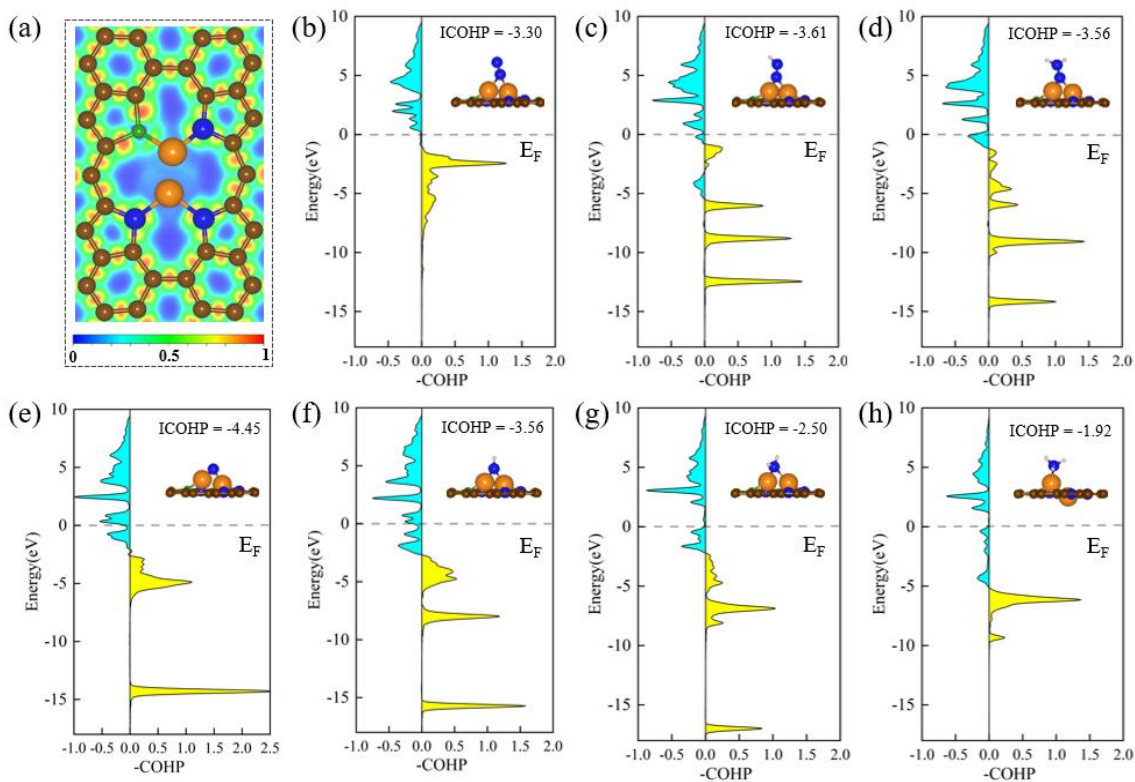

**Figure 3.** (a) The electron localization function (ELF) of Fe$_2$N$_3$B@G catalyst, the color represents the degree of electron localization, where the red area represents highly localized electron density and the blue area represents dispersed electron distribution. (b-h) The projected crystal orbital Hamiltonian population (COHP) between different reaction intermediates (*N$_2$, *NNH, *NNH$_2$, *N, *NH, *NH$_2$, and *NH$_3$) and Fe atoms in the distal pathway of the NRR for Fe$_2$N$_3$B@G catalyst, the cyan and yellow areas represent the contributions of antibonding and bonding states, respectively.

In the process of catalyst design, the regulation of electronic structure plays a crucial role. Figure 3 (a) shows the electron localization function (ELF) of Fe$_2$N$_3$B@G catalyst, revealing the significant effect of doping B atoms on the electronic structure of the catalyst and its important role in nitrogen activation. The blue area near the Fe atom indicates that its electron distribution is moderately localized. This localized characteristic enhances the bonding between Fe atoms and N$_2$ molecules and provides the necessary electron supply for breaking the N≡N triple bond. In contrast, the prominent red area near B atom shows a highly localized electron density. This phenomenon originates from the optimization of the local electronic environment by moderate electron redistribution during the bonding process between the B, Fe and C atoms. The introduction of the B atom not only enhances the stability of the Fe-C and Fe-B bonds, but also promotes the bonding between Fe and nitrogen molecules by regulating the electron distribution around the Fe atom, thereby further enhancing the electronic activity of the dual-Fe atom. This localization effect provides a more stable and efficient electron supply environment for the dual-Fe atom in the catalytic process, significantly enhancing the adsorption and activation ability of the catalyst for nitrogen molecules. In addition, the doping of B atoms also enhances the overall structural stability of the catalyst by regulating the electron distribution, providing important support for the efficient NRR. These results show that B-N co-doping is a key design strategy for optimizing catalyst performance. By doping B atoms in appropriate amounts, the activity and efficiency of the catalyst in the NRR can be significantly improved, providing a theoretical basis for further optimizing catalyst design.

Figure 3 (b-h) shows the projected crystal orbital Hamiltonian population (COHP) between the Fe atom and the reaction intermediate in the distal pathway of the NRR on the $Fe_2N_3B@G$ catalyst. COHP analysis reveals the dynamic changes in the bonding and antibonding contributions of the Fe-N bond during the reduction process, where the yellow area represents the bonding electron density and the cyan area represents the antibonding electron density. From the initial adsorption of $N_2$ to the final desorption of $NH_3$, the bonding strength of the Fe-N bond shows a dynamic regulation characteristic, which contributes to the efficient catalytic performance of the catalyst. In the $N_2$ stage (see Figure 3(b)), Fe atoms form a preliminary interaction with $N_2$ molecules and show good electron donation ability (ICOHP = -3.30), laying the foundation for the activation of the N≡N triple bond. The electron distribution in this stage shows that $N_2$ molecules can be effectively adsorbed on the catalyst surface and achieve initial weakening of the triple bond through electron transfer, thus creating favorable conditions for subsequent reaction steps. As the reaction enters the NNH and $NNH_2$ stages (see Figure 3(c-d)), the bonding contribution is significantly enhanced, indicating that Fe atoms effectively provide electron support in these two stages, thereby further promoting the activation and hydrogenation processes of nitrogen molecules. The strong bonding force in these stages provides a guarantee for the stability of the intermediates and the smooth progress of subsequent reactions. In the N stage (Figure 3(e)), the bonding contribution reaches a peak value (ICOHP = -4.45), indicating that the single nitrogen atom has the strongest bonding force on the Fe surface. This stage marks the complete breakage of the N≡N triple bond. In the NH stage (Figure 3(f), ICOHP = -3.56) and the $NH_2$ stage (Figure 3(g), ICOHP = -2.50), the bonding contribution gradually weakens and the anti-bonding effect gradually increases, indicating that with the increase in the degree of hydrogenation, the intermediate gradually detaches from the Fe surface. In the $NH_3$ stage (Figure 3(h), ICOHP = -1.92), the bonding strength of the Fe-N bond is further reduced, and the $NH_3$ molecule can be desorbed from the catalyst surface, thereby completing the entire nitrogen reduction process. Therefore, the $Fe_2N_3B@G$ catalyst successfully achieves efficient activation of $N_2$ molecules and stable adsorption of intermediates by dynamically regulating the strength of the Fe-N bond, while ensuring the release of the final product $NH_3$. The dynamic adaptability to electronic behavior constitutes the linchpin of the $Fe_2N_3B@G$'s exceptional efficacy and furnishes a pivotal theoretical underpinning for the continued refinement of catalyst architecture.Figure 4 shows the electron transfer behavior and electrostatic potential distribution of the $Fe_2N_3B@G$ catalyst in the NRR, revealing the synergistic effect of the electron behavior of different regions during the reaction and its influence on the intermediates. Figure 4(a) defines three key regions of the catalyst: Moiety-1 represents the different NRR intermediates in the reaction, Moiety-2 is the Fe dual-atom site, which is the main electron supply and intermediate activation region; Moiety-3 represents the B and N co-doped region and the graphene network around it, which mainly plays a role in regulating the local electronic environment and enhancing the stability of the catalyst structure. Figure 4(b) shows the trend of charge changes in these three regions during the NRR. The electron transfer of Moiety-2 shows obvious fluctuations during the reaction, especially in the stage from N to $NH_3$, indicating that the dual-Fe site effectively activates the nitrogen molecule and promotes the hydrogenation process of the intermediate by providing electron support. Moiety-3 shows a relatively stable charge change, indicating that its main role is to assist the function of the dual-Fe site by stabilizing the electronic environment. Moiety-1 reflects the electron accumulation during the reaction, especially in the N stage, where the electron change reaches the maximum value, indicating the breaking of N≡N triple bond of $N_2$ molecule. Figure S3 further supports the analysis by showing charge variations in Moiety-1, Moiety-2, and Moiety-3 for $Fe_2N_2B_2@G$, $Fe_2NB_3@G$, and $Fe_2B_4@G$ catalysts. The charge fluctuations in Moiety-2 highlight the role of dual-Fe site in electron transfer during nitrogen activation and hydrogenation. Moiety-3 shows stable charge behavior, indicating its role in stabilizing the catalyst's electronic environment, while Moiety-1 reflects electron accumulation, especially during the $N_2$ activation.

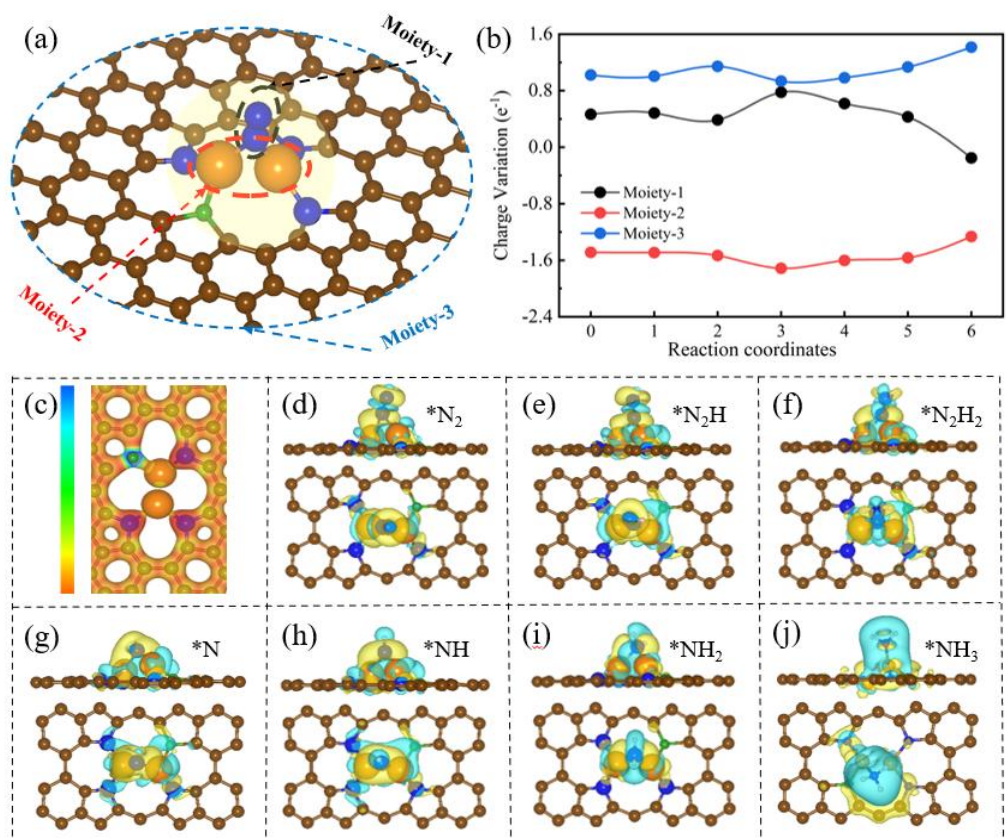

**Figure 4.** Electronic behavior and electrostatic potential distribution of distal pathway of Fe$_2$N$_3$B@G catalyst in NRR. (a) Regional division of catalyst, Moiety-1 represents different reaction intermediates (*N$_2$, *NNH, *NNH$_2$, *N, *NH, *NH$_2$, and *NH$_3$), Moiety-2 is dual-Fe site, and Moiety-3 represents the doped B and N region and its surrounding graphene network. (b) Changes in electron transfer. The horizontal axis corresponds to the different reaction stages in Figure 4 (d-j), showing the charge changes in different regions. (c) Electrostatic potential (ESP) diagram of Fe$_2$N$_3$B@G catalyst, showing the enrichment of electrons at dual-Fe site. (d-j) Electron density distribution of different reaction intermediates on the catalyst surface, the yellow area indicates electron density accumulation, the cyan area indicates electron density removal, and the isosurface threshold is 0.001$e$ per Bohr³.

Figure 4(c) shows the electrostatic potential (ESP) distribution of the Fe$_2$N$_3$B@G catalyst, showing that the dual-Fe site (Moiety-2) is enriched with electrons during the reaction, while the B and N co-doped region (Moiety-3) shows a higher positive electrostatic potential. This electrostatic potential distribution further proves the function of the Fe dual-atom site as an electron supply center. Meanwhile, the B and N co-doped regions enhance the overall stability of the catalyst and the adsorption capacity of intermediates by adjusting the electronic environment. Figure 4(d-j) shows the electron density distribution of the Fe$_2$N$_3$B@G catalyst during the NRR and the electronic interaction between the dual-Fe site and the nitrogen intermediate. In the early stage of the reaction, the dual-Fe site interacts with the N$_2$ molecule through electron supply, promoting the breaking of the N≡N triple bond and activating the nitrogen molecule. As the reaction progresses, Fe atoms continuously provide electronic support and enhance the bonding electron density of the Fe-N bond. Especially in the N stage, the adsorption capacity of Fe atoms for nitrogen reaches its maximum, marking the key step of nitrogen reduction for the complete breaking of the N≡N bond. Subsequently, during the hydrogenation of nitrogen molecules, the strength of the Fe-N bond gradually weakened, and the electron density gradually dispersed, reflecting that the adsorption force of Fe atoms on intermediates gradually decreased, but remained stable. Finally, in the $NH_3$ stage, the bonding electron density of the Fe-N bond was significantly weakened, and the ammonia molecule was able to desorb to complete the NRR. Overall, the $Fe_2N_3B$@G catalyst achieves efficient $N_2$ activation, hydrogenation and final ammonia molecule desorption by dynamically adjusting the electron density of the Fe-N bond, demonstrating the physical mechanism of regulating of electronic supply in $N_2$ activation process for achieving the catalyst's high efficiency performance. The regulation of electronic behavior in this process provides an important theoretical basis for optimizing catalyst design.

3.4 Stability and catalytic selectivity of $Fe_2N_3B$@G atomic catalyst toward NRR

The stability of a catalyst is crucial to its practical application, especially the thermal stability and structural stability under different reaction conditions. Hence, we combined the formation energy and thermal stability of the $Fe_2N_3B$@G catalyst to explore the stability dependence on different numbers of Fe atoms, as seen in Figure 5 . Figure 5(a) shows the top and side views of the Fe atoms gradually increasing from 1 to 2, 3 and 4. As the number of Fe atoms increases, the interaction between Fe atoms gradually increases, and the coordination environment on the catalyst surface is changed, affecting the structural stability of the catalyst. Figure 5(b) shows the trend of the formation energy of the $Fe_2N_3B$@G catalyst with the Fe chemical potential. As seen in Figure 5(b), $Fe_1N_3B$@G (black line) exhibits a lower formation energy at a lower Fe chemical potential, indicating that it has strong stability under low Fe concentration conditions. As the number of Fe atoms increases to 2 ($Fe_2N_3B$@G, red line), its formation energy decreases significantly, and under higher Fe concentration conditions, the formation energy of $Fe_2N_3B$@G is confirmed to be the lowest, indicating that $Fe_2N_3B$@G has the strongest stability at higher Fe concentrations. This shows that $Fe_2N_3B$@G maintain high structural stability at different Fe concentrations and effectively enhance the stability of the dual-Fe site, thereby improving the catalytic performance. Further increasing the number of Fe atoms ($Fe_3N_3B$@G, blue line and $Fe_4N_3B$@G, green line), the formation energy of Fe$_x$N$_3$B@G increases significantly, indicating that the catalyst is less stable at higher Fe concentrations. Therefore, higher Fe concentrations enhance the interaction between Fe atoms, which reduces the stability of the catalyst and may cause Fe atoms to agglomerate or migrate, thus reducing the catalytic performance.

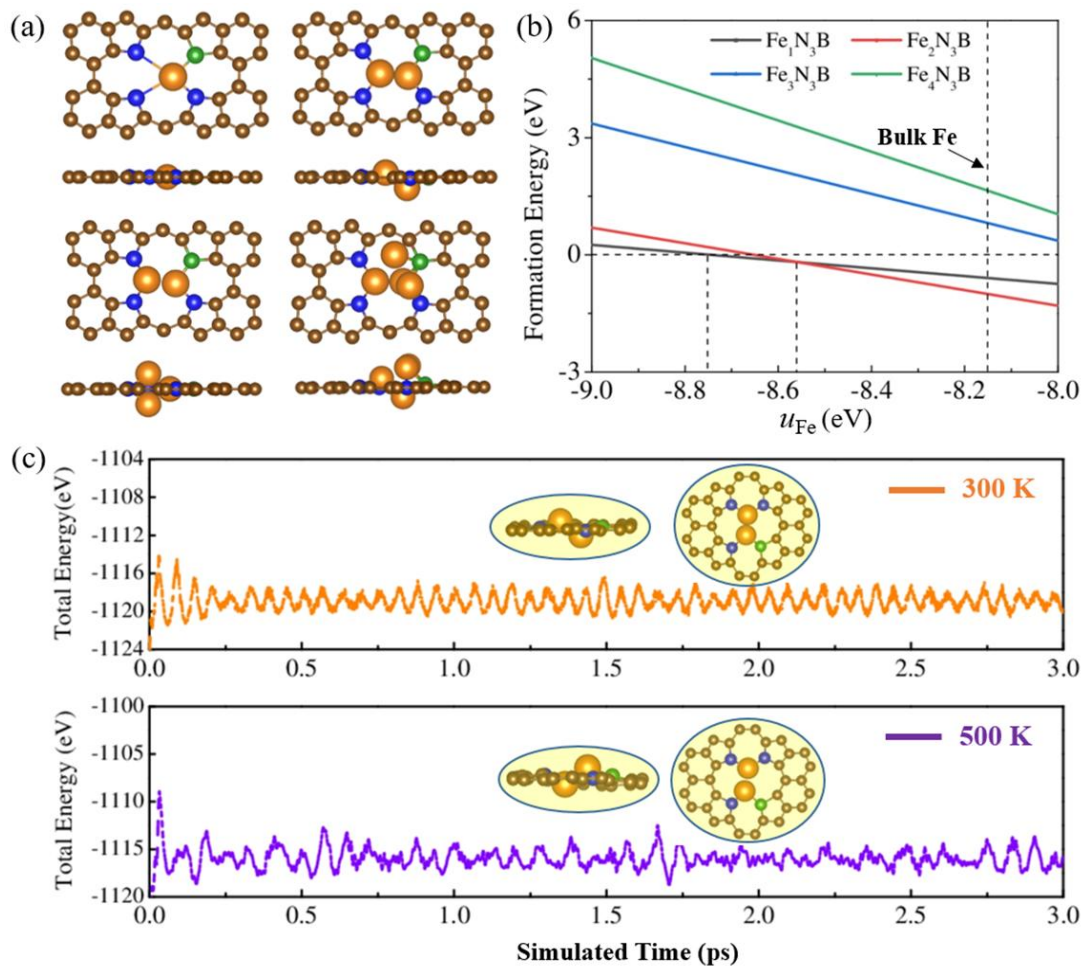

**Figure 5.** (a) Stability analysis of Fe$_x$N$_3$B@G catalyst under different Fe atoms. (a) Top view and side views of Fe$_x$N$_3$B@G catalyst with the number of Fe atoms gradually increasing from 1 to 4. (b) The trend of the formation energy of Fe$_x$N$_3$B@G catalyst changing with Fe chemical potential, with the horizontal axis being Fe chemical potential and the vertical axis being formation energy. (c) Molecular dynamics simulation results of Fe$_2$N$_3$B@G at 300 K and 500 K for time 3 ps.

The molecular dynamics simulation in Figure 5(c) further reveals the thermal stability of Fe$_2$N$_3$B@G at 300 K and 500 K. At both temperatures, the basic structural framework of Fe$_2$N$_3$B@G can be well preserved intact and no obvious Fe-N or Fe-B chemical bonds fracture phenomena are observed at temperature up to 500 K, which clearly demonstrate the high-temperature resistance property. The good adaptation to catalytic reactions under different temperature conditions further enhance Fe$_2$N$_3$B@G potential as a catalyst in practical applications.

The selectivity issue in NRR has always been one of the core challenges in catalyst design, especially in the competitive HER for hydrogen generation. In order to effectively inhibit the HER and improve the selectivity of NRR, we further explore the performance of $Fe_2N_3B@G$ catalyst in HER through free energy analysis. Figure 6 shows the free energy change of $Fe_xN_3B@G$ catalyst in HER and the process of kinetic hydrogen generation on $Fe_2N_3B@G$ catalyst. The datas in Figure 6(a) show that there are significant differences in the free energy of the HER under different B-doping environments. The free energy of the $Fe_2N_3B@G$ (black line) catalyst in the HER is -1.45 eV, indicating that its hydrogen evolution performance is poor and its overpotential is high, which effectively avoids competition with the NRR. In contrast, $Fe_2B_4@G$ (green line) exhibits the lowest hydrogen evolution free energy (-1.13 eV), but it is still high, indicating that its hydrogen evolution performance is poor and its overpotential is high, hence it can effectively inhibit the HER and promote the NRR. The free energies of $Fe_2N_2B_2@G$ (-1.46 eV) and $Fe_2NB_3@G$ (-1.56 eV) in HER are higher, which further enhance their ability to avoid the HER, thereby providing favorable conditions for the NRR.

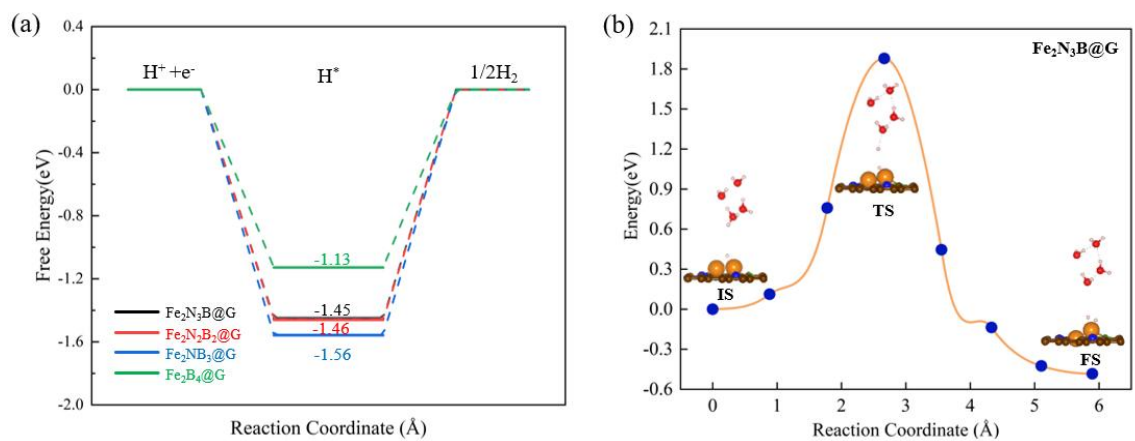

**Figure 6.** (a) Free energy curves of HER on $Fe_2N_3B$, $Fe_2N_2B_2$, $Fe_2NB_3$, and $Fe_2B_4$ catalysts. (b) Minimum energy pathway for $H_2$ generation on $Fe_2N_3B@G$ catalyst.

Figure 6(b) shows the simulated results of hydrogen addition and $H_2$ generation of the $Fe_2N_3B@G$ catalyst. Using the minimum energy path method, the figure shows the energy changes in the hydrogenation process, from the initial state (IS) to the transition state (TS) and then to the final state (FS). The energy peak at the transition state (TS) is 1.95 eV, indicating that the generation of $H_2$ molecules requires to overcome a high energy barrier, making the hydrogen generation process more difficult on the surface of the $Fe_2N_3B@G$ catalyst. This result shows that the $Fe_2NB_3@G$ catalyst can effectively inhibit the HER, thereby reducing the occurrence of competitive reactions. In Figure S4, the various steps in the NRR show that the highest energy barrier is just 0.33 eV, significantly lower than that for the HER. This indicates that the NRR process on Fe$_2$N$_3$B@G is highly favorable and the catalyst can effectively promote NRR over HER, confirming its high NRR selectivity and catalytic efficiency.

3.5 Study on machine learning molecular dynamics simulations of Fe$_2$N$_3$B@G atomic catalyst toward NRR

To gain a deeper understanding of the reaction mechanism of Fe$_2$N$_3$B@G catalyst, this work combines MLMD simulation to explore the dynamic behavior of the catalyst in NRR. Figure 7 shows the MLMD simulation results of Fe$_2$N$_3$B@G in NRR, covering the key processes from N$_2$ activation to the final generation of NH$_3$. Figure 7 (a-b) show the energy and force comparison of the hydrogenation reaction of N$_2$ to NNH, where the comparison of the energy calculated by the machine learning model with the DFT results. The N$_2$ adsorption and hydrogenation process of the Fe$_2$N$_3$B@G catalyst has an extremely low root mean square error (RMSE), with an energy error of 0.001 eV/atom and a force error of 0.132 eV/Å, indicating that the NRR performance of the catalyst can be accurately predicted by MLMD simulation. In the early stage of N$_2$ adsorption, the dual-Fe site of Fe$_2$N$_3$B@G provides the necessary electronic support, effectively promoting the breaking of the N≡N triple bond and the activation of nitrogen, providing a favorable reaction environment for the subsequent hydrogenation step. Figure 7 (c-g) show the process of N$_2$ being hydrogenated to NNH observed during the MLMD simulation. In this process, the Fe$_2$N$_3$B@G catalyst provides electronic support through the dual-Fe site, allowing the N$_2$ molecule to effectively break and form NNH on the catalyst surface. The simulation results of this process further verified the superior performance of the Fe$_2$N$_3$B@G catalyst. The dual-Fe site promoted the activation of nitrogen molecules by precisely regulating the electron supply. Figure 7 (j-n) show the process of NH$_2$ being hydrogenated to generate NH$_3$. In this process, the Fe$_2$N$_3$B@G catalyst still exhibited superior catalytic performance, ensuring the smooth generation and desorption of ammonia molecules by continuously providing electron support. These two key steps play an important role in the NRR, and the Fe$_2$N$_3$B@G catalyst ensures the stable adsorption and desorption of nitrogen and ammonia molecules through efficient catalysis. The MLMD results further proved that the Fe$_2$N$_3$B@G catalyst can not only effectively activate nitrogen, but also ensure the efficient generation and desorption of ammonia on dual-Fe site. Its excellent nitrogen activation and ammonia generation capabilities provide a theoretical basis for catalyst design and optimization, and demonstrate the broad application prospects of Fe$_2$N$_3$B@G in NRRs.

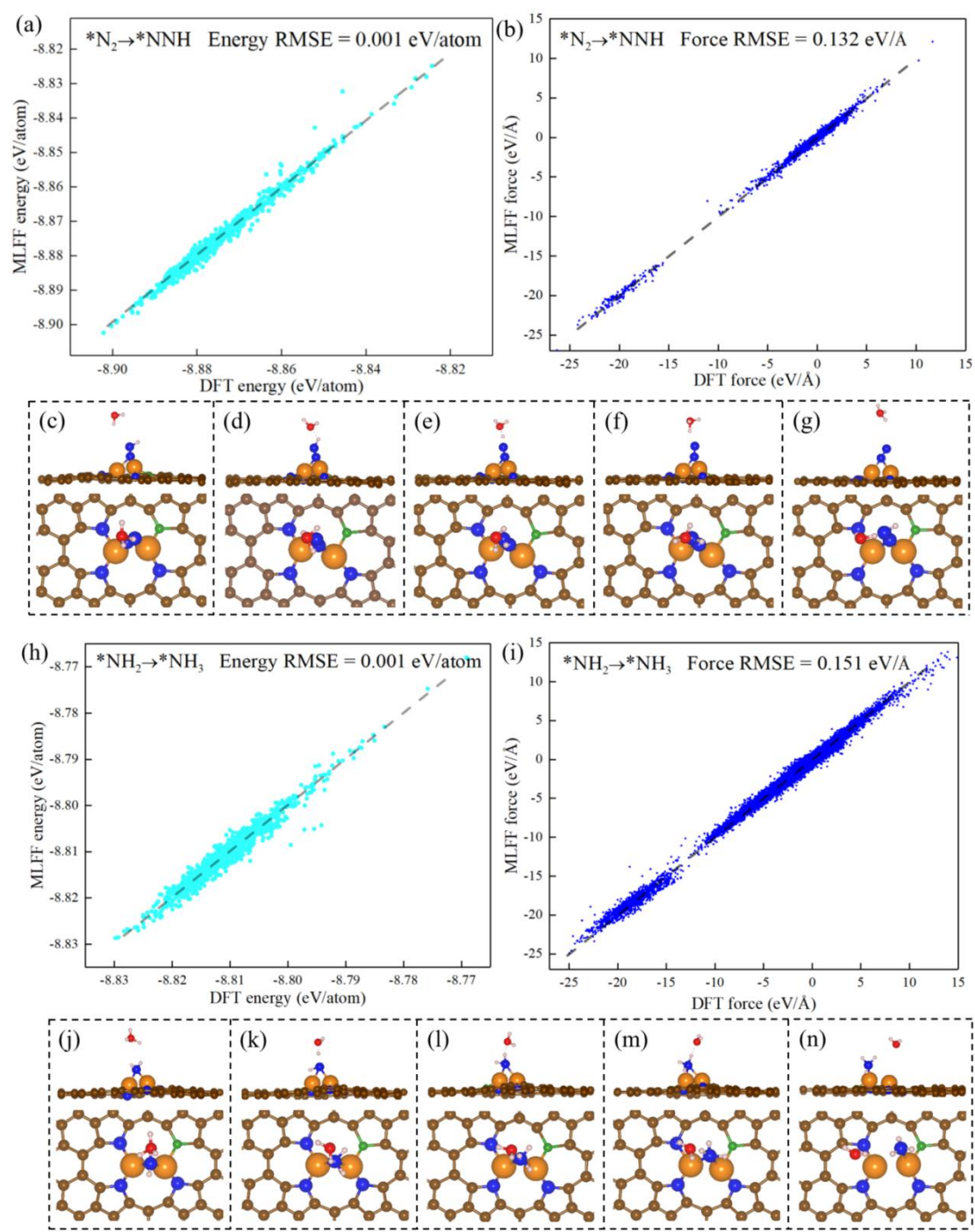

**Figure 7.** MLMD simulation results of NRR on Fe$_2$N$_3$B@G catalyst. (a-b) Comparison of energy and force of *N$_2$ to *NNH hydrogenation reaction with DFT calculation results. (c-g) Top view and side view of MLMD of *N$_2$ to *NNH hydrogenation process. (h) Comparison of energy and force of *NH$_2$ to *NH$_3$ hydrogenation reaction with DFT calculation results. (j-n) Top view and side view of MLMD of *NH$_2$ to *NH$_3$ hydrogenation process.

## 4. Conclusions

In summary, this work innovatively designed a Fe$_2$N$_3$B@G catalyst with excellent NRR performance by modulating the structural and electronic properties of dual-Fe atomic site via $B_xN_y$ co-doping strategy on graphene. $Fe_2N_3B@G$ catalyst showed excellent performance in the adsorption and activation of $N_2$ molecules and $NH_3$ generation through precise electron supply, showing lower free energy of 0.32 eV on NRR distal pathway. The computed electron localization function, crystal orbital Hamiltonian population, and electrostatic potential map have collectively unveiled the profound insights into the enhanced NRR kinetics exhibited by the $Fe_2N_3B@G$ catalyst. This enhancement is attributed to the meticulous optimization of the Fe-Fe electronic environment, modulation of Fe-N bond strength, and persistent electronic support facilitated by the strategic $N_3B$ co-doping. This sophisticated interplay of electronic factors ensures efficient $N_2$ breakage and hydrogenation processes. In addition, the catalyst has a higher hydrogen evolution overpotential, and effectively avoids the competition with the NRR, further enhancing the selectivity toward NRR. In particular, the MD simulations up to 500K and the advanced MLMD simulations were utilized to rigorously validate the stability and exceptional activity of the $Fe_2N_3B@G$ catalyst within the entire NRR pathway, demonstrating its great potential in practical applications. In summary, the design of $Fe_2N_3B@G$ catalyst not only provides new ideas for efficient NRR catalysis, but also provides strong theoretical support for the structural and electronic optimization of dual-atom atomic catalysts.

**Author contributions**
**Zhe Zhang:** analyzed the data, wrote the paper. **Wen Ma:** performed the simulations, analyzed the data. **Jiajie Qia:** analyzed the data. **Xiaoliang Wu:** analyzed the data. **Shaowen Yu:** analyzed the data. **Weiye Hou:** analyzed the data. **Xiang Huang:** analyzed the data. **Rubin Huo:** performed the simulations, analyzed the data. **Hongbo Wu:** analyzed the data, wrote the paper. **Yusong Tu:** guided the research, analyzed the data.

**Conflicts of interest**
There are no conflicts to declare.

**Acknowledgments**
This work was funded by the National Natural Science Foundation of China (Nos. 12347169, 12404087), the Natural Science Foundation of Jiangsu Province (No. BK20240892), the Natural Science Foundation of Jiangsu Higher Education Institutions

**References**

1. B. H. Suryanto, K. Matuszek, J. Choi, R. Y. Hodgetts, H.-L. Du, J. M. Bakker, C. S. Kang, P. V. Cherepanov, A. N. Simonov and D. R. MacFarlane, *Science*, 2021, **372**, 1187-1191.
2. A. N. Singh, R. Anand, M. Zafari, M. Ha and K. S. Kim, *Adv. Energy Mater.*, 2024, **14**, 2304106.
3. G. Hai and H. Wang, *Small Methods*, 2023, **7**, e2300756.
4. D. R. MacFarlane, P. V. Cherepanov, J. Choi, B. H. R. Suryanto, R. Y. Hodgetts, J. M. Bakker, F. M. Ferrero Vallana and A. N. Simonov, *Joule*, 2020, **4**, 1186-1205.
5. M. Yang, J. Yang, N. He, S. Wang, H. Ni, J. Yuan, Y. Kang, Y. Liu, C. Zhou, L. Tong, B. Lu, X. Liu, Q. Wang, S. Huang, B. Feng, G. Guo, S. Han and Z. Han, *ACS Appl. Nano Mater.*, 2024, **8**, 179-188.
6. J. Sun, D. Zheng, F. Deng, S. Liu, Y. Xie, Y. Liu, J. Xu and W. Liu, *Appl. Surf. Sci.*, 2024, **644**, 158802.
7. X. Mao, X. Bai, G. Wu, Q. Qin, A. P. O'Mullane, Y. Jiao and A. Du, *J. Am. Chem. Soc.*, 2024, **146**, 18743-18752.
8. Y. Wu, C. He and W. Zhang, *J. Am. Chem. Soc.*, 2022, **144**, 9344-9353.
9. F. Guo, J. Ma, X. Deng and H. Gao, *Int. J. Hydrogen Energy*, 2024, **74**, 183-192.
10. T. Bo, S. Cao, N. Mu, R. Xu, Y. Liu and W. Zhou, *Appl. Surf. Sci.*, 2023, **612**, 155916.
11. N. Sathishkumar and H. T. Chen, *ACS Appl Mater Interfaces*, 2023, **15**, 15545−15560.
12. Y. Zhao, Y. Jiang, Y. Mo, Y. Zhai, J. Liu, A. C. Strzelecki, X. Guo and C. Shan, *Small*, 2023, **19**, e2207240.
13. R. Zhao, Y. Chen, H. Xiang, Y. Guan, C. Yang, Q. Zhang, Y. Li, Y. Cong and X. Li, *ACS Appl Mater Interfaces*, 2023, **15**, 6797-6806.
14. P. Shu, X. Qi, Q. Peng, Y. Chen, X. Gong, Y. Zhang, F. Ouyang and Z. Sun, *Mol. Catal.*, 2023, **539**.
15. N. Zhang, M.-Y. Wang and J.-Y. Liu, *Vacuum*, 2023, **210**.
16. Z. Zhang, B. Zheng, H. Tian, Y. He, X. Huang, S. Ali and H. Xu, *Phys. Chem. Chem. Phys.*, 2022, **24**, 18265-18271.
17. L. Hu, Z. Zhang, B.-B. Zheng and J.-Z. Zhao, *Surf. Interfaces*, 2023, **39**, 102916.
18. Z. L. Zhao, S. Yang, S. Wang, Z. Zhang, L. Zhao, Q. Wang and X. Zhang, *Advanced Science*, 2024, 2411705.
19. F. Li and Q. Tang, *Nanoscale*, 2019, **11**, 18769-18778.
20. W. Yang, H. Huang, X. Ding, Z. Ding, C. Wu, I. D. Gates and Z. Gao, *Electrochim. Acta*, 2020, **335**, 135667.

21. S. Jia, H. Zhu, R. Cao, Q. Wu, C. Wu, Q. Zhou, P. Liu, B. Li, A. Li and Y. Li, *Int. J. Hydrogen Energy*, 2024, **83**, 367-377.
22. X. Guo, J. Gu, S. Lin, S. Zhang, Z. Chen and S. Huang, *J. Am. Chem. Soc.*, 2020, **142**, 5709-5721.
23. R. Hu, Y. Li, Q. Zeng, F. Wang and J. Shang, *ChemSusChem*, 2020, **13**, 3636-3644.
24. T. Deng, C. Cen, H. Shen, S. Wang, J. Guo, S. Cai and M. Deng, *J Phys Chem Lett*, 2020, **11**, 6320-6329.
25. Z. Zhang, X. Huang and H. Xu, *ACS Appl. Mater. Interfaces*, 2021, **13**, 43632−43640.
26. Q.-P. Zhao, W.-X. Shi, J. Zhang, Z.-Y. Tian, Z.-M. Zhang, P. Zhang, Y. Wang, S.-Z. Qiao and T.-B. Lu, *Nature Synth.*, 2024, **3**, 1078
27. L. Jiao and L. Guo, *Inorg. Chem.*, 2022, **61**, 18574-18589.
28. Y. Xu, W. Li, L. Chen, W. Li, W. Feng and X. Qiu, *Inorg. Chem.*, 2023, **62**, 5253-5261.
29. N. Sathishkumar and H.-T. Chen, *J. Phys. Chem. C*, 2023, **127**, 994-1005.
30. G. Kresse and J. Furthmuller, *Phys. Rev. B*, 1996, **54**, 11169-11186.
31. J. P. Perdew, M. Ernzerhof and K. Burke, *J. Chem. Phys.*, 1996, **105**, 9982-9985.
32. P. E. Blöchl, *Phys. Rev. B*, 1994, **50**, 17953-17979.
33. G. Kresse and D. Joubert, *Phys. Rev. B*, 1999, **59**, 1758-1775.
34. B. Hammer, L. B. Hansen and J. K. Norskov, *Phys. Rev. B*, 1999, **59**, 7413-7421.
35. G. Henkelman, B. P. Uberuaga and H. Jónsson, *J. Chem. Phys.*, 2000, **113**, 9901-9904.
36. R. Jinnouchi, F. Karsai and G. Kresse, *Phys. Rev. B*, 2019, **100**, 014105
37. J. S. Smith, O. Isayev and A. E. Roitberg, *Chem Sci*, 2017, **8**, 3192-3203.
38. C. Wang, A. Tharval and J. R. Kitchin, *Mol. Simulat.*, 2018, **44**, 623-630.
39. N. Yao, X. Chen, Z. H. Fu and Q. Zhang, *Chem. Rev.*, 2022, **122**, 10970-11021.
40. R. Jinnouchi, J. Lahnsteiner, F. Karsai, G. Kresse and M. Bokdam, *Phys. Rev. Lett.*, 2019, **122**, 225701.